\documentclass[%
 reprint,
superscriptaddress,
 amsmath,amssymb,
 aps,
]{revtex4-1}

\usepackage{graphicx}
\usepackage{dcolumn}
\usepackage{bm}

\usepackage[utf8]{inputenc}
\usepackage[T1]{fontenc}
\usepackage{mathptmx}
\usepackage{etoolbox}
\usepackage[dvipsnames]{xcolor}
\usepackage{siunitx}
\usepackage{float}

\begin{document}

\preprint{APS/123-QED}

\title{Integrated phononic waveguide on thin-film lithium niobate on diamond}

\author{Sultan Malik}
\thanks{These authors contributed equally}
\author{Felix M. Mayor}
\thanks{These authors contributed equally}
\author{Wentao Jiang}
\affiliation{Department of Applied Physics and Ginzton Laboratory, Stanford University, 348 Via Pueblo Mall, Stanford, California 94305, USA}

\author{Hyunseok Oh}
\author{Carl Padgett}
\author{Viraj Dharod}
\author{Jayameenakshi Venkatraman}

\author{Ania C. Bleszynski Jayich}
\affiliation{Department of Physics, University of California, Santa Barbara, California 93106, USA}
\author{Amir H. Safavi-Naeini}%
\email{safavi@stanford.edu}
\affiliation{Department of Applied Physics and Ginzton Laboratory, Stanford University, 348 Via Pueblo Mall, Stanford, California 94305, USA}

\date{\today}

\begin{abstract}
We demonstrate wavelength-scale phononic waveguides formed by transfer-printed thin-film lithium niobate (LN) on bulk diamond (LNOD), a material stack that combines the strong piezoelectricity of LN with the high acoustic velocity and color-center compatibility of diamond. We characterize a delay line based on a $\SI{100}{\micro\meter}$ long phononic waveguide at room and cryogenic temperatures. The total insertion loss through the device at \SI{4}{\kelvin} is $\SI{-5.8}{\deci\bel}$, corresponding to a $> \SI{50}{\percent}$ transducer efficiency, at a frequency of $\SI{2.8}{\giga\hertz}$. Our work represents a step towards phonon-mediated hybrid quantum systems consisting of strain-sensitive color centers in diamond. 
\end{abstract}

\maketitle


Acoustic waves travel orders of magnitude more slowly than electromagnetic waves, making on-chip acoustic devices particularly effective in RF signal processing applications where they are widely used as delay lines~\cite{lu20195} and filters~\cite{zhang2020surface}. Traditionally, surface acoustic wave (SAW)~\cite{xu2025thin,shao2019phononic} and bulk acoustic wave (BAW)~\cite{bi2008bulk,chu2017quantum} devices have been employed for both classical and quantum applications. Reducing the size of these devices could lead to higher component density, more complex device architectures, as well as lower cross talk and greater confinement of excitation energy. One way to achieve this is through the use of \textit{phononic} integrated circuits. Analogous to photonic integrated circuits~\cite{thomson2016roadmap} that guide and manipulate light on a chip, phononic integrated circuits~\cite{fu2019phononic,mayor2021gigahertz} offer similar capabilities for sound by using wavelength-scale acoustic waveguides. Such guiding can be achieved by patterning a thin-film material that has a lower propagation velocity than the underlying substrate. As in photonics, a larger velocity mismatch leads to greater modal confinement. To drive and read out these guided modes electrically, we need a piezoelectric transducer that converts signals between the electric and acoustic domains. Several platforms have been developed using a top piezoelectric thin-film layer consisting of (GaN)~\cite{fu2019phononic}, lithium niobate (LN)~\cite{mayor2021gigahertz}, aluminum nitride (AlN)~\cite{ding2024integrated} or aluminum scandium nitride (AlScN)~\cite{deng2025strongly} placed on top a substrate with high acoustic velocity.

Diamond is particularly attractive as a substrate material because of its high acoustic velocity ($>\SI[per-mode=symbol]{12}{\kilo\meter\per\second}$) and the ability to host highly coherent optically addressable color centers~\cite{atature_material_2018}. Among these, negatively charged silicon vacancy centers (SiV) in diamond~\cite{meesala2018strain, maity2020coherent} are especially promising for quantum networking and memory applications~\cite{sukachev2017silicon}. They also exhibit a very large strain susceptibility (up to $\sim \SI{100}{\tera\hertz}$), allowing efficient coupling to acoustic fields and facilitating spin-phonon based quantum information protocols~\cite{lemonde2018phonon}. Guided wave phononic circuits are an attractive way for ``wiring up'' defect centers in diamond using acoustic fields. By combining diamond with highly piezoelectric thin-film lithium niobate, smaller, broader-band, and lower-loss transducers that operate at cryogenic temperatures~\cite{yamamoto2023low} become possible. Integrating thin-film LN on diamond requires bonding high-quality LN onto a SiV-containing diamond substrate. While SAWs in LN on diamond have recently been demonstrated~\cite{xu2025thin}, wavelength-scale acoustic waveguides have yet to be realized in this material platform. 

In this work, we demonstrate efficient driving of a \SI{100}{\micro\meter}-long wavelength-scale phononic waveguide in LN on bulk diamond. We make use of the large piezoelectric coupling $k_\text{eff}^2 \sim \SI{21}{\percent}$ achievable in LN to design a compact ($ \sim 2 \times\SI{22}{\micro\meter}^2$) interdigital transducer (IDT) to excite \SI{2.8}{\giga\hertz} acoustic waves in the waveguide. Our fabrication approach involves transfer printing patterned thin-film LN on a different substrate material onto a diamond substrate. Finally, we characterize the device scattering parameters at room and cryogenic temperatures ($T \approx \SI{4}{\kelvin}$) and estimate spin-phonon coupling in these acoustic waveguides.

\begin{figure*}[!htbp]
\includegraphics[scale=1]{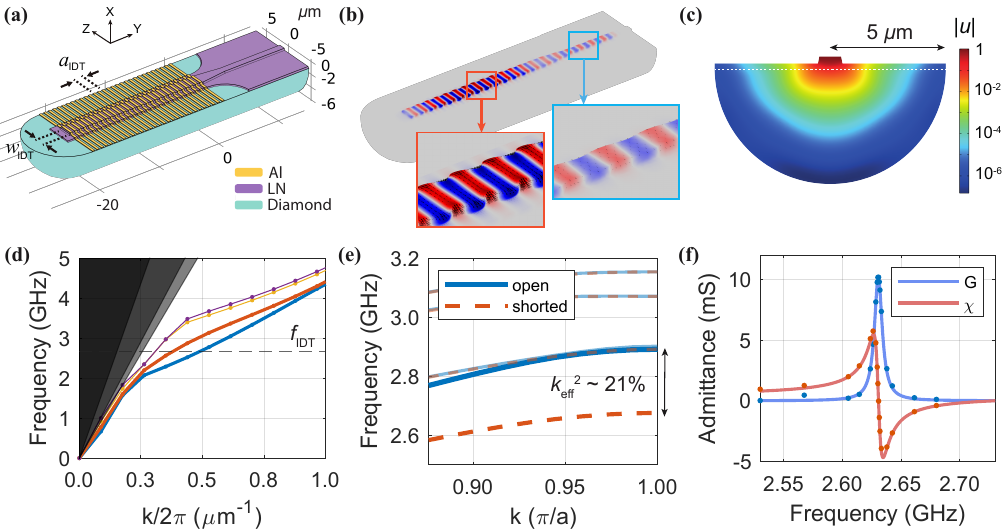}
\caption{\label{fig:sim} 
\textbf{Design of the mechanical waveguide and piezoelectric transducer.} 
\textbf{(a)} Finite-element model of one end of the device showing the IDT, taper, and waveguide regions.
\textbf{(b)} Simulated driven mechanical mode profile of the IDT-taper-waveguide model at its resonant frequency with colors representing the transverse horizontal component of the displacement field. 
\textbf{(c)} Normalized displacement field of the quasi-Love guided mode in the rib waveguide at $\SI{2.67}{\giga\hertz}$. The white dashed line represents the LN-diamond boundary.
\textbf{(d)} Mechanical bands of the LN waveguide. The blue (red) band corresponds to the quasi-Love (quasi-Rayleigh) mode. Black cones represent the bulk acoustic and surface acoustic waves in diamond.
\textbf{(e)} First four mechanical bands of the IDT unit cell near the \textit{X} point. The difference in the open (solid blue line) and the shorted (dashed red line) electrode boundary conditions highlights the strongly-coupled mode with $k_\text{eff}^2 \sim \SI{21}{\percent}$.
\textbf{(f)} Admittance response from the frequency-domain simulations. Solid lines are fits to the simulated data.
}
\end{figure*}

A key challenge in phononic circuits is achieving efficient and broadband transduction of electrical energy into the desired propagating mechanical mode in the waveguide. Maximizing this efficiency, in turn, requires impedance matching between the $\SI{50}{\Omega}$ transmission line and the transducer, and a good mechanical mode match between the transducer mode and the phononic waveguide mode of interest. The choice of LN, with its high piezoelectric and high dielectric coefficients, enables impedance matching with a relatively small total transducer area. To maximize the piezoelectric coupling of the IDT, we orient the IDT on $X$-cut LN such that the electric field points along the $Y$ direction of the crystal, as shown in Fig.~\ref{fig:sim}(a), and couples to the $YZ$ strain component, via the $d_{24} \sim \SI[per-mode=symbol]{70}{\pico\coulomb\per\newton}$ component of the piezoelectric tensor, transducing a shear horizontal (SH) mode in the IDT region as shown in Fig.~\ref{fig:sim}(b). We choose the IDT width to be $w_\text{IDT}  = \SI{2}{\micro\meter}$, on the same order as the waveguide width of $w_\text{}  = \SI{1}{\micro\meter}$, enabling us to use a simple linear taper to connect the two regions (see Fig.~\ref{fig:sim}(a)), and also have efficient mode matching~\cite{mayor2021gigahertz, dahmani2020piezoelectric}. We remove the LN (purple) slab layer around the IDT region to minimize parasitic excitation of surface and bulk acoustic waves.

To first identify the frequency of the guided acoustic wave that can be driven in the rib waveguide, we use finite-element method to perform quasi-two-dimensional simulations of the waveguide cross section using anisotropic diamond as the substrate. For a waveguide of thickness $t = \SI{300}{\nano\meter}$, width $w_\text{}  = \SI{1}{\micro\meter}$, and a slab layer of thickness $t_\text{s} = \SI{200}{\nano\meter}$, we calculate the mechanical dispersion diagram of the waveguide, plotted in Fig.~\ref{fig:sim}(d). The longitudinal and shear bulk acoustic waves, and the surface acoustic waves in diamond form the boundaries of the shaded area, and modes above these lines generally form a continuum of acoustic radiation modes. The two lowest frequency bands correspond to the quasi-Love and quasi-Rayleigh modes, respectively. To avoid potential hybridization between these two modes, we choose the quasi-Love mode as the mode of interest in the waveguide at $\SI{2.67}{\giga\hertz}$ because it has maximum separation in $k$ from the quasi-Rayleigh mode around that frequency. We show the tight confinement of this quasi-Love mode by plotting the normalized displacement $|u|$ of the waveguide cross section at $\SI{2.67}{\giga\hertz}$ in Fig.~\ref{fig:sim}(c), where the displacement $\SI{5}{\micro\meter}$ away from the center of the waveguide is five orders of magnitude smaller.

To excite the quasi-Love mode at $\SI{2.67}{\giga\hertz}$ in the waveguide, we choose an IDT periodicity $a_\text{IDT}  = \SI{1.5}{\micro\meter}$ and an electrode duty cycle of $\SI{45}{\percent}$. We simulate the IDT unit cell with periodic boundary condition and plot the first four bands near the $X$ point with open (solid blue line) and shorted (dashed red line) boundary conditions on the electrodes as shown in Fig.~\ref{fig:sim}(e). We identify the strongly-coupled mode at $\SI{2.67}{\giga\hertz}$ by noting its large frequency shift (highlighted as thicker lines) with the changing electric circuit boundary condition from open to closed. From this shift, we estimate~\cite{dahmani2020piezoelectric} the piezoelectric coupling to be $k_\text{eff}^2 \sim \SI{21}{\percent}$. Solving the model to obtain the admittance at low frequencies, we obtain the dielectric capacitance $C_\text{0} = \SI{4.3}{\femto\farad}$ per unit cell. This allows us to calculate the required area of the IDT using the formula~\cite{sarabalis_s-band_2020, dahmani2020piezoelectric}
\begin{eqnarray} \label{eq1}
    A=\frac{\pi^2}{8} \frac{G_\text{0}}{\omega_\text{IDT}^2 c_\text{0}} \frac{\gamma}{k_{\mathrm{eff}}^2},
\end{eqnarray}
where $c_\text{0}$ is the capacitance per unit area, and the peak conductance $G_\text{0} = \SI{20}{\milli\siemens}$ and the bandwidth $\gamma /2\pi = \SI{50}{\mega\hertz}$ are assumed to match the IDT to a \SI{50}{\ohm} transmission line over a relatively broad bandwidth. Using $\omega_\text{IDT} / 2\pi = \SI{2.67}{\giga\hertz}$, we obtain the estimate of the required number of unit cells to be $N=15$. From the full simulation of the IDT-taper-waveguide region driven at its resonant frequency, we verify the SH motion in both the IDT and the waveguide regions as shown in Fig.~\ref{fig:sim}(b). The simulated admittance is shown in Fig.~\ref{fig:sim}(f) with $G_\text{0} = \SI{10.2}{\milli\siemens}$ and $\gamma /2\pi = \SI{8.14}{\mega\hertz}$. Additionally, we estimate that $\sim \SI{77}{\percent}$ of the mechanical power generated by the IDT is guided in the waveguide, with $\sim \SI{11}{\percent}$ going in the slab and the remaining in the diamond substrate. This sets the maximum efficiency of the IDT, with the actual efficiency reduced by factors such as resistive loss of the metal electrodes, material loss of LN, and excess acoustic scattering from roughness and nonuniformity.

\begin{figure}[!htbp]
\includegraphics[scale=1]{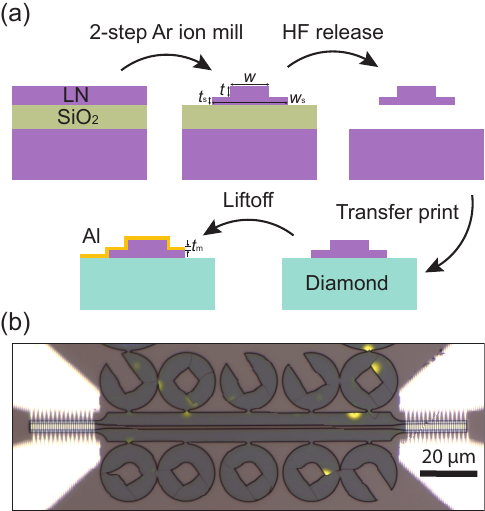}
\caption{\label{fig:fab}  \textbf{Device fabrication.} \textbf{(a)} Fabrication process of the lithium niobate on diamond integrated phononic waveguide. The LN rib waveguide is defined on a separate chip before being transferred to a diamond substrate. The aluminum IDT is patterned subsequently via liftoff. \textbf{(b)} Optical microscope image of the device. The circular structures attached to the LN slab are release anchors that broke off during transfer printing.}
\end{figure}


In contrast to other piezoelectric materials such as aluminum nitride~\cite{ding2024integrated}, it is challenging to sputter or epitaxially grow high-quality single-crystal LN. Therefore, material platforms such as thin-film lithium niobate on insulator (LNOI) are typically produced via bulk crystal growth followed by bonding and then Smart Cut or thinning via grinding/polishing. To heterogeneously integrate thin-film LN on diamond (LNOD), we start with already prepared thin-film material and use a transfer printing technique~\cite{meitl2006transfer,jiang2023optically} to place it onto diamond. The simplified fabrication process is illustrated in Fig.~\ref{fig:fab}(a). We start with a LNOI piece (commercially available from NanoLN) with \SI{500}{\nano\meter} thick X-cut LN on \SI{2}{\micro\meter} thick silica sitting on a \SI{500}{\micro\meter} thick LN substrate. Next, we define the $w=\SI{1}{\micro\meter}$ wide rib waveguide with e-beam lithography in hydrogen silsesquioxane (HSQ) resist and etch $t = \SI{300}{\nano\meter}$ into the LN with Ar-ion milling. To be able to undercut the waveguide for transfer printing, we repeat the same process to etch the remaining $t_\text{s} = \SI{200}{\nano\meter}$ LN slab. Thanks to the high confinement in the waveguide, we only need to keep a $w_\text{s}  = \SI{10}{\micro\meter}$ wide LN slab around the waveguide. Near the IDT, we only keep $w_\text{s}  = \SI{4}{\micro\meter}$ to reduce excitation of unwanted bulk and surface acoustic waves. The patterned LNOI sample is then cleaned in a piranha solution together with the target (100) diamond substrate (from Element Six, electronic grade, $\SI{2}{\milli\meter}\times\SI{2}{\milli\meter}$, thickness: \SI{0.5}{\milli\meter}). Before transfer printing, the LN is further cleaned and simultaneously released from the silica in \SI{5}{\percent} hydrofluoric acid (HF) at \SI{43}{\celsius}. We use polydimethylsiloxane (PDMS) for transfer printing the LN onto the diamond sample, and follow the same procedure as in~\cite{jiang2023optically}. Once the LN has been transferred, the LNOD is annealed in air at \SI{400}{\celsius} for \SI{8}{\hour}. Finally, the IDTs are defined with aligned e-beam lithography in CSAR 62.0013 resist, followed by an angled evaporation of $t_\text{m}=\SI{100}{\nano\meter}$ thick aluminum and liftoff in N-methyl-2-pyrrolidone (NMP). The angled evaporation helps the aluminum electrodes climb onto the LN from the diamond. An optical microscope image of the final device is shown in Fig.~\ref{fig:fab}(b). The fabricated device has a waveguide length $L=\SI{100}{\micro\meter}$, an IDT period $a_\text{IDT}=\SI{1.5}{\micro\meter}$ and $N = 15$ periods on each IDT.

\begin{figure}[!htbp]
\includegraphics[scale=1]{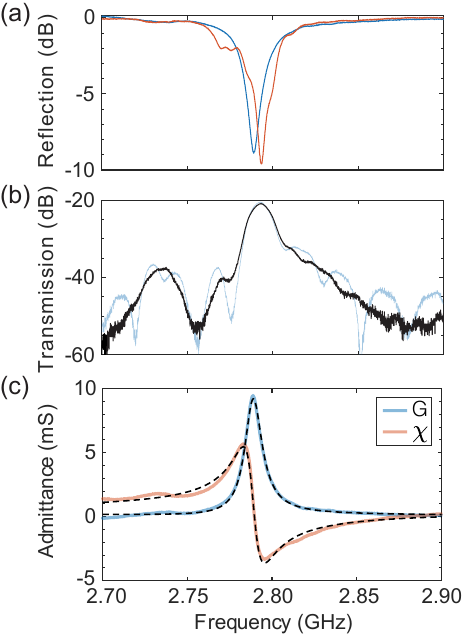}
\caption{\label{fig:roomT} \textbf{Room Temperature characterization.} \textbf{(a)} Microwave reflection with the $S$-parameter $|S_{11}|$ ($|S_{22}|$) shown in blue (red). \textbf{(b)} Microwave transmission $|S_{21}|$ with the cable response removed (blue). The black line shows the response with the microwave crosstalk filtered out. \textbf{(c)} Admittance measurement extracted from $S_{11}$ showing the conductance $G$ (blue) and the susceptance $\chi$ (red). The dashed lines are a Lorentzian fit.}
\end{figure}

\begin{figure}[!htbp]
\includegraphics[scale=1]{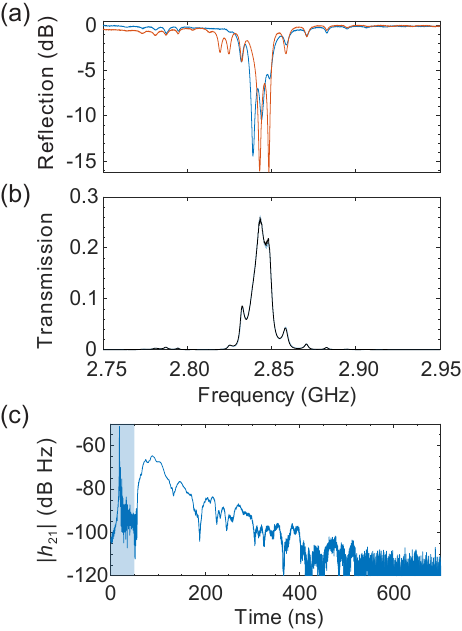}
\caption{\label{fig:lowT} \textbf{Cryogenic characterization at \SI{4}{\kelvin}.} \textbf{(a)} Microwave reflection $|S_{11}|$ ($|S_{22}|$) shown in blue (red). \textbf{(b)} Microwave transmission $|S_{21}|$ (raw data: blue, filtered response: black). The peak $|S_{21}|^2$ is \SI{26}{\percent} (\SI{-5.8}{\deci\bel}). \textbf{(c)} Impulse response $|h_{21}|$ in transmission. The shaded area highlights the microwave crosstalk which can be filtered out.}
\end{figure}

To characterize the device, we first mount the sample on a printed circuit board (PCB) using GE varnish and then mount the PCB inside a Montana Instruments cryostat. The IDTs on both ends of the acoustic waveguide are wirebonded to the PCB allowing us to measure the two-port scattering parameters $S_{ij}$ of the device using a vector network analyzer (VNA). Before wirebonding to the device, we measured $S_{ij}^\text{c}$ of the cables connected to the bare PCB. The termination is open, resulting in reflection of the microwaves. We use this data to remove the cable response from the device $S$-parameters with $S_{ii} = S_{ii}^\text{raw}/S_{ii}^\text{c}$. The calibrated room temperature microwave reflection $|S_{11}|^2$ ($|S_{22}|^2$) for port 1 (2) is shown in Fig.~\ref{fig:roomT}(a). For both IDTs, there is a dip in the reflection spectrum at $\sim \SI{2.79}{\giga\hertz}$ where the microwaves are transduced into the acoustic domain. Similarly, when we look at the transmission $|S_{21}|^2$ (see Fig.~\ref{fig:roomT}(b)), we see a peak at \SI{2.79}{\giga\hertz}. The peak transmission of $|S_{21}|^2 \sim \SI{-21}{\deci\bel}$ is comparable to the room temperature values measured for related integrated waveguide devices at gigahertz frequencies~\cite{mayor2021gigahertz,ding2024integrated}. Here, we have removed the cable response through $S_{21} = S_{21}^\text{raw}/\sqrt{S_{11}^\text{c}S_{22}^\text{c}}$. From an inverse Fourier transform of $S_{21}$, we obtain the impulse response $h_{21}$ and use it to filter out the microwave crosstalk. The impulse response for a different data set at \SI{4}{\kelvin} is shown in Fig.~\ref{fig:lowT}(c). This microwave crosstalk happens on a much faster time scale than the delay $\tau_\text{d}$ from the acoustic wave propagating through the waveguide and we can filter it out by simply setting $h_{21}(\tau \ll \tau_\text{d})=0$. The filtered response is shown as the black curve in Fig.~\ref{fig:roomT}(b). Before wirebonding, and using a fully calibrated microwave probe station, we previously measured the delay between the two IDTs for a device with a \SI{100}{\micro\meter} and one with a \SI{145}{\micro\meter} long waveguide. From this, we obtain a group velocity in the waveguide of $v_\text{g,meas} \approx \SI{2.8 \pm 0.3}{\kilo\meter\per\second}$, in agreement with the simulated value ($v_\text{g,sim} = \SI{2.84}{\kilo\meter\per\second}$). Finally, we extract the admittance of the IDT from $S_{11}$ and plot the conductance $G$ and susceptance $\chi$ (Fig.~\ref{fig:roomT}(c)). With a Lorentzian fit of the complex admittance, we obtain a peak conductance $G_0 = \SI{9.2}{\milli\siemens}$, bandwidth $\gamma/2\pi = \SI{12.5}{\mega\hertz}$ and center frequency $\omega_\text{IDT}/2\pi = \SI{2.788}{\giga\hertz}$, in reasonable agreement with the simulations.

Operation at cryogenic temperatures is necessary for experiments using color centers~\cite{sukachev2017silicon} which could be incorporated in the diamond substrate of phononic waveguides in the future. Cooling down the device has the added advantage that acoustic and resistive losses are significantly reduced~\cite{shao2019phononic,fu2019phononic,mayor2021gigahertz}, helping increase transmission through the device. We therefore cool down the device to a temperature of $T\approx \SI{4}{\kelvin}$ and measure its scattering parameters. The cable response was measured in a separate cooldown and is removed from the data. The microwave input power at the IDT is $\sim \SI{1}{\micro\watt}$ and we show the calibrated reflection $|S_{11}|^2$ and $|S_{22}|^2$ in Fig.~\ref{fig:lowT}(a). Compared to room temperature, the transducer center frequency is blue-shifted by \SI{50}{\mega\hertz} to \SI{2.84}{\giga\hertz}. In Fig.~\ref{fig:lowT}(b), we show the measured microwave transmission $|S_{21}|^2$ through the device with a peak of \SI{26}{\percent} (\SI{-5.8}{\deci\bel}), a significant reduction in insertion loss compared to room temperature. Additionally, we observe features separated by a free spectral range (FSR) of $\sim\SI{12}{\mega\hertz}$ that we attribute to standing waves forming in the waveguide due to reflection between the two IDTs. The lower acoustic loss at cryogenic temperatures allows these standing waves to be more prominent than at room temperature. While it is challenging to assign a precise cavity length, this FSR roughly matches what we expect for the group velocity of $v_\text{g} \approx \SI[per-mode=symbol]{3}{\kilo\meter\per\second}$ found using the impulse response at room temperature and a waveguide length of \SI{100}{\micro\meter}.

While we have shown successful transfer printing of patterned $\SI{100}{\micro\meter}$ straight LN waveguides on diamond, the yield for bigger and more complicated patterned structures, such as longer straight waveguides and ring resonators, remains low. Improvement in LN pattern and tether design around the release anchors is expected to improve the yield of such devices. Improved yield of structures like longer straight waveguides of varying length and ring resonators~\cite{fu2019phononic,mayor2021gigahertz} would allow us to collect better statistics and accurate measurements of propagation loss in the waveguides. Alternatively, yield may also be improved by adopting a direct bonding approach where a thin-film of LN is bonded on a diamond substrate, followed by direct patterning of LN on diamond, thereby eliminating the need for transfer printing. While this would allow fabrication of larger and more complex devices, there is a potential risk of damaging the diamond surface and the implanted color centers due to the etching process.


In conclusion, we have introduced and demonstrated a platform for LN-based wavelength-scale integrated phononic waveguides on diamond. The combination of the large piezoelectric coefficient of LN, resulting in strong electromechanical coupling, and the high speed of sound in diamond, resulting in tight confinement of the mechanical mode, has allowed us to demonstrate a transducer with insertion loss less than \SI{3}{\deci\bel}, corresponding to a $> \SI{50}{\percent}$ transducer efficiency at $\SI{4}{\kelvin}$. 

Given the high efficiency of our transducer, our platform provides an attractive route toward a spin-phonon interface for the SiV color center in diamond, which hosts a spin qubit with a large strain susceptibility of up to $d=\SI{100}{\tera\hertz}$/strain ~\cite{maity2020coherent, meesala2018strain} and a splitting that can be tuned with a magnetic field into the GHz range ~\cite{hepp_electronic_2014, maity2020coherent}. The guided acoustic mode generates strain in diamond and can evanescently couple ~\cite{ding2024integrated} to an SiV spin below the diamond surface. From the waveguide simulation shown in Fig.~\ref{fig:sim}(c), we can extract the largest zero-point strain component $\epsilon_\text{zpf}$ at $\sim \SI{100}{\nano\meter}$ depth, allowing us to estimate a spin-phonon coupling rate of $g_\text{sp}/2\pi = d \times \epsilon_\text{zpf} \sim \SI{24}{\kilo\hertz}$/phonon, on the same order as the one estimated in ~\cite{ding2024integrated, xu2025thin}. From this, we estimate a single spin-phonon cooperativity $C_\text{sp} = 4g_\text{sp}^2/\gamma_\text{s}\gamma_\text{p}$, where $\gamma_\text{s}$ is the spin dephasing rate and $\gamma_\text{p}$ is the mechanical mode's energy dissipation rate. Assuming a reasonable $\gamma_\text{s}/2\pi \sim \SI{0.016}{\mega\hertz}$ (corresponding to a $T^*_\text{2} \sim \SI{10}{\micro\second}$ for isotopically purified diamond) at $\SI{}{\milli\kelvin}$ temperature ~\cite{sukachev_silicon-vacancy_2017} and estimating an effective $\gamma_\text{p} \sim 1/t_\text{wg} \sim v_\text{g}/L\sim 2\pi \times\SI{4.6}{\mega\hertz}$ for the waveguide, we get a single-spin cooperativity of $C_\text{sp} \sim\SI{3.2e-2}{}$. The possibility of fabricating a ring resonator structure on such a platform, as discussed earlier, and assuming a quality factor of at least $\sim\SI{5e4}{}$ ~\cite{mayor2021gigahertz}, would increase the single-spin cooperativity to $C_\text{sp} \sim\SI{2.6}{}$. This would allow a path toward coherent acoustic control of spins, potentially enabling phonon-mediated hybrid quantum systems, such as on-chip networking of multiple spins and phononic links between superconducting qubits and spin-based quantum memories.

\section*{Acknowledgments}
The authors would like to thank Kevin~K.S.~Multani and Matthew~P.~ Maksymowych for fabrication assistance, and Samuel Gyger and Andre~G.~Primo for useful discussions. This work was primarily supported by the US Army Research Office (ARO)/Laboratory for Physical Sciences (LPS) Modular Quantum Gates (ModQ) program (Grant No. W911NF-23-1-0254). Some of this work was funded by the US Department of Energy through grant no.\ DE-AC02-76SF00515 and via the Q-NEXT Center, the NSF QLCI program through grant number OMA-2016245, and by the National Science Foundation CAREER award no.\ ECCS-1941826. 
We also thank Amazon Web Services Inc.\ for their financial support.
Device fabrication was performed at the Stanford Nano Shared Facilities (SNSF) and the Stanford Nanofabrication Facility (SNF), supported by NSF award ECCS-2026822.


\section*{Competing interests}
A.H.S.-N. is an Amazon Scholar.

\section*{Data availability}
The data that support the findings of this study are available from the corresponding author upon reasonable request.


\bibliography{ref}

\begin{thebibliography}{23}%
\makeatletter
\providecommand \@ifxundefined [1]{%
 \@ifx{#1\undefined}
}%
\providecommand \@ifnum [1]{%
 \ifnum #1\expandafter \@firstoftwo
 \else \expandafter \@secondoftwo
 \fi
}%
\providecommand \@ifx [1]{%
 \ifx #1\expandafter \@firstoftwo
 \else \expandafter \@secondoftwo
 \fi
}%
\providecommand \natexlab [1]{#1}%
\providecommand \enquote  [1]{``#1''}%
\providecommand \bibnamefont  [1]{#1}%
\providecommand \bibfnamefont [1]{#1}%
\providecommand \citenamefont [1]{#1}%
\providecommand \href@noop [0]{\@secondoftwo}%
\providecommand \href [0]{\begingroup \@sanitize@url \@href}%
\providecommand \@href[1]{\@@startlink{#1}\@@href}%
\providecommand \@@href[1]{\endgroup#1\@@endlink}%
\providecommand \@sanitize@url [0]{\catcode `\\12\catcode `\$12\catcode
  `\&12\catcode `\#12\catcode `\^12\catcode `\_12\catcode `\%12\relax}%
\providecommand \@@startlink[1]{}%
\providecommand \@@endlink[0]{}%
\providecommand \url  [0]{\begingroup\@sanitize@url \@url }%
\providecommand \@url [1]{\endgroup\@href {#1}{\urlprefix }}%
\providecommand \urlprefix  [0]{URL }%
\providecommand \Eprint [0]{\href }%
\providecommand \doibase [0]{http://dx.doi.org/}%
\providecommand \selectlanguage [0]{\@gobble}%
\providecommand \bibinfo  [0]{\@secondoftwo}%
\providecommand \bibfield  [0]{\@secondoftwo}%
\providecommand \translation [1]{[#1]}%
\providecommand \BibitemOpen [0]{}%
\providecommand \bibitemStop [0]{}%
\providecommand \bibitemNoStop [0]{.\EOS\space}%
\providecommand \EOS [0]{\spacefactor3000\relax}%
\providecommand \BibitemShut  [1]{\csname bibitem#1\endcsname}%
\let\auto@bib@innerbib\@empty
\bibitem [{\citenamefont {Lu}\ \emph {et~al.}(2019)\citenamefont {Lu},
  \citenamefont {Yang}, \citenamefont {Li}, \citenamefont {Breen},\ and\
  \citenamefont {Gong}}]{lu20195}%
  \BibitemOpen
  \bibfield  {author} {\bibinfo {author} {\bibfnamefont {R.}~\bibnamefont
  {Lu}}, \bibinfo {author} {\bibfnamefont {Y.}~\bibnamefont {Yang}}, \bibinfo
  {author} {\bibfnamefont {M.-H.}\ \bibnamefont {Li}}, \bibinfo {author}
  {\bibfnamefont {M.}~\bibnamefont {Breen}}, \ and\ \bibinfo {author}
  {\bibfnamefont {S.}~\bibnamefont {Gong}},\ }\href@noop {} {\bibfield
  {journal} {\bibinfo  {journal} {IEEE Transactions on Microwave Theory and
  Techniques}\ }\textbf {\bibinfo {volume} {68}},\ \bibinfo {pages} {573}
  (\bibinfo {year} {2019})}\BibitemShut {NoStop}%
\bibitem [{\citenamefont {Zhang}\ \emph {et~al.}(2020)\citenamefont {Zhang},
  \citenamefont {Lu}, \citenamefont {Zhou}, \citenamefont {Link}, \citenamefont
  {Yang}, \citenamefont {Li}, \citenamefont {Huang}, \citenamefont {Ou},\ and\
  \citenamefont {Gong}}]{zhang2020surface}%
  \BibitemOpen
  \bibfield  {author} {\bibinfo {author} {\bibfnamefont {S.}~\bibnamefont
  {Zhang}}, \bibinfo {author} {\bibfnamefont {R.}~\bibnamefont {Lu}}, \bibinfo
  {author} {\bibfnamefont {H.}~\bibnamefont {Zhou}}, \bibinfo {author}
  {\bibfnamefont {S.}~\bibnamefont {Link}}, \bibinfo {author} {\bibfnamefont
  {Y.}~\bibnamefont {Yang}}, \bibinfo {author} {\bibfnamefont {Z.}~\bibnamefont
  {Li}}, \bibinfo {author} {\bibfnamefont {K.}~\bibnamefont {Huang}}, \bibinfo
  {author} {\bibfnamefont {X.}~\bibnamefont {Ou}}, \ and\ \bibinfo {author}
  {\bibfnamefont {S.}~\bibnamefont {Gong}},\ }\href@noop {} {\bibfield
  {journal} {\bibinfo  {journal} {IEEE Transactions on Microwave Theory and
  Techniques}\ }\textbf {\bibinfo {volume} {68}},\ \bibinfo {pages} {3653}
  (\bibinfo {year} {2020})}\BibitemShut {NoStop}%
\bibitem [{\citenamefont {Xu}\ \emph {et~al.}(2025)\citenamefont {Xu},
  \citenamefont {Ding}, \citenamefont {Cornell}, \citenamefont {Mohideen},
  \citenamefont {Yeh}, \citenamefont {Kuruma}, \citenamefont {Magalhaes},
  \citenamefont {Shams-Ansari}, \citenamefont {Pingault},\ and\ \citenamefont
  {Loncar}}]{xu2025thin}%
  \BibitemOpen
  \bibfield  {author} {\bibinfo {author} {\bibfnamefont {Z.}~\bibnamefont
  {Xu}}, \bibinfo {author} {\bibfnamefont {S.~W.}\ \bibnamefont {Ding}},
  \bibinfo {author} {\bibfnamefont {E.}~\bibnamefont {Cornell}}, \bibinfo
  {author} {\bibfnamefont {S.}~\bibnamefont {Mohideen}}, \bibinfo {author}
  {\bibfnamefont {M.}~\bibnamefont {Yeh}}, \bibinfo {author} {\bibfnamefont
  {K.}~\bibnamefont {Kuruma}}, \bibinfo {author} {\bibfnamefont
  {L.}~\bibnamefont {Magalhaes}}, \bibinfo {author} {\bibfnamefont
  {A.}~\bibnamefont {Shams-Ansari}}, \bibinfo {author} {\bibfnamefont
  {B.}~\bibnamefont {Pingault}}, \ and\ \bibinfo {author} {\bibfnamefont
  {M.}~\bibnamefont {Loncar}},\ }\href@noop {} {\bibfield  {journal} {\bibinfo
  {journal} {arXiv preprint arXiv:2505.08895}\ } (\bibinfo {year}
  {2025})}\BibitemShut {NoStop}%
\bibitem [{\citenamefont {Shao}\ \emph {et~al.}(2019)\citenamefont {Shao},
  \citenamefont {Maity}, \citenamefont {Zheng}, \citenamefont {Wu},
  \citenamefont {Shams-Ansari}, \citenamefont {Sohn}, \citenamefont {Puma},
  \citenamefont {Gadalla}, \citenamefont {Zhang}, \citenamefont {Wang} \emph
  {et~al.}}]{shao2019phononic}%
  \BibitemOpen
  \bibfield  {author} {\bibinfo {author} {\bibfnamefont {L.}~\bibnamefont
  {Shao}}, \bibinfo {author} {\bibfnamefont {S.}~\bibnamefont {Maity}},
  \bibinfo {author} {\bibfnamefont {L.}~\bibnamefont {Zheng}}, \bibinfo
  {author} {\bibfnamefont {L.}~\bibnamefont {Wu}}, \bibinfo {author}
  {\bibfnamefont {A.}~\bibnamefont {Shams-Ansari}}, \bibinfo {author}
  {\bibfnamefont {Y.-I.}\ \bibnamefont {Sohn}}, \bibinfo {author}
  {\bibfnamefont {E.}~\bibnamefont {Puma}}, \bibinfo {author} {\bibfnamefont
  {M.}~\bibnamefont {Gadalla}}, \bibinfo {author} {\bibfnamefont
  {M.}~\bibnamefont {Zhang}}, \bibinfo {author} {\bibfnamefont
  {C.}~\bibnamefont {Wang}},  \emph {et~al.},\ }\href@noop {} {\bibfield
  {journal} {\bibinfo  {journal} {Physical Review Applied}\ }\textbf {\bibinfo
  {volume} {12}},\ \bibinfo {pages} {014022} (\bibinfo {year}
  {2019})}\BibitemShut {NoStop}%
\bibitem [{\citenamefont {Bi}\ and\ \citenamefont {Barber}(2008)}]{bi2008bulk}%
  \BibitemOpen
  \bibfield  {author} {\bibinfo {author} {\bibfnamefont {F.~Z.}\ \bibnamefont
  {Bi}}\ and\ \bibinfo {author} {\bibfnamefont {B.~P.}\ \bibnamefont
  {Barber}},\ }\href@noop {} {\bibfield  {journal} {\bibinfo  {journal} {IEEE
  microwave magazine}\ }\textbf {\bibinfo {volume} {9}},\ \bibinfo {pages} {65}
  (\bibinfo {year} {2008})}\BibitemShut {NoStop}%
\bibitem [{\citenamefont {Chu}\ \emph {et~al.}(2017)\citenamefont {Chu},
  \citenamefont {Kharel}, \citenamefont {Renninger}, \citenamefont {Burkhart},
  \citenamefont {Frunzio}, \citenamefont {Rakich},\ and\ \citenamefont
  {Schoelkopf}}]{chu2017quantum}%
  \BibitemOpen
  \bibfield  {author} {\bibinfo {author} {\bibfnamefont {Y.}~\bibnamefont
  {Chu}}, \bibinfo {author} {\bibfnamefont {P.}~\bibnamefont {Kharel}},
  \bibinfo {author} {\bibfnamefont {W.~H.}\ \bibnamefont {Renninger}}, \bibinfo
  {author} {\bibfnamefont {L.~D.}\ \bibnamefont {Burkhart}}, \bibinfo {author}
  {\bibfnamefont {L.}~\bibnamefont {Frunzio}}, \bibinfo {author} {\bibfnamefont
  {P.~T.}\ \bibnamefont {Rakich}}, \ and\ \bibinfo {author} {\bibfnamefont
  {R.~J.}\ \bibnamefont {Schoelkopf}},\ }\href@noop {} {\bibfield  {journal}
  {\bibinfo  {journal} {Science}\ }\textbf {\bibinfo {volume} {358}},\ \bibinfo
  {pages} {199} (\bibinfo {year} {2017})}\BibitemShut {NoStop}%
\bibitem [{\citenamefont {Thomson}\ \emph {et~al.}(2016)\citenamefont
  {Thomson}, \citenamefont {Zilkie}, \citenamefont {Bowers}, \citenamefont
  {Komljenovic}, \citenamefont {Reed}, \citenamefont {Vivien}, \citenamefont
  {Marris-Morini}, \citenamefont {Cassan}, \citenamefont {Virot}, \citenamefont
  {F{\'e}d{\'e}li} \emph {et~al.}}]{thomson2016roadmap}%
  \BibitemOpen
  \bibfield  {author} {\bibinfo {author} {\bibfnamefont {D.}~\bibnamefont
  {Thomson}}, \bibinfo {author} {\bibfnamefont {A.}~\bibnamefont {Zilkie}},
  \bibinfo {author} {\bibfnamefont {J.~E.}\ \bibnamefont {Bowers}}, \bibinfo
  {author} {\bibfnamefont {T.}~\bibnamefont {Komljenovic}}, \bibinfo {author}
  {\bibfnamefont {G.~T.}\ \bibnamefont {Reed}}, \bibinfo {author}
  {\bibfnamefont {L.}~\bibnamefont {Vivien}}, \bibinfo {author} {\bibfnamefont
  {D.}~\bibnamefont {Marris-Morini}}, \bibinfo {author} {\bibfnamefont
  {E.}~\bibnamefont {Cassan}}, \bibinfo {author} {\bibfnamefont
  {L.}~\bibnamefont {Virot}}, \bibinfo {author} {\bibfnamefont {J.-M.}\
  \bibnamefont {F{\'e}d{\'e}li}},  \emph {et~al.},\ }\href@noop {} {\bibfield
  {journal} {\bibinfo  {journal} {Journal of Optics}\ }\textbf {\bibinfo
  {volume} {18}},\ \bibinfo {pages} {073003} (\bibinfo {year}
  {2016})}\BibitemShut {NoStop}%
\bibitem [{\citenamefont {Fu}\ \emph {et~al.}(2019)\citenamefont {Fu},
  \citenamefont {Shen}, \citenamefont {Xu}, \citenamefont {Zou}, \citenamefont
  {Cheng}, \citenamefont {Han},\ and\ \citenamefont {Tang}}]{fu2019phononic}%
  \BibitemOpen
  \bibfield  {author} {\bibinfo {author} {\bibfnamefont {W.}~\bibnamefont
  {Fu}}, \bibinfo {author} {\bibfnamefont {Z.}~\bibnamefont {Shen}}, \bibinfo
  {author} {\bibfnamefont {Y.}~\bibnamefont {Xu}}, \bibinfo {author}
  {\bibfnamefont {C.-L.}\ \bibnamefont {Zou}}, \bibinfo {author} {\bibfnamefont
  {R.}~\bibnamefont {Cheng}}, \bibinfo {author} {\bibfnamefont
  {X.}~\bibnamefont {Han}}, \ and\ \bibinfo {author} {\bibfnamefont {H.~X.}\
  \bibnamefont {Tang}},\ }\href@noop {} {\bibfield  {journal} {\bibinfo
  {journal} {Nature Communications}\ }\textbf {\bibinfo {volume} {10}},\
  \bibinfo {pages} {2743} (\bibinfo {year} {2019})}\BibitemShut {NoStop}%
\bibitem [{\citenamefont {Mayor}\ \emph {et~al.}(2021)\citenamefont {Mayor},
  \citenamefont {Jiang}, \citenamefont {Sarabalis}, \citenamefont {McKenna},
  \citenamefont {Witmer},\ and\ \citenamefont
  {Safavi-Naeini}}]{mayor2021gigahertz}%
  \BibitemOpen
  \bibfield  {author} {\bibinfo {author} {\bibfnamefont {F.~M.}\ \bibnamefont
  {Mayor}}, \bibinfo {author} {\bibfnamefont {W.}~\bibnamefont {Jiang}},
  \bibinfo {author} {\bibfnamefont {C.~J.}\ \bibnamefont {Sarabalis}}, \bibinfo
  {author} {\bibfnamefont {T.~P.}\ \bibnamefont {McKenna}}, \bibinfo {author}
  {\bibfnamefont {J.~D.}\ \bibnamefont {Witmer}}, \ and\ \bibinfo {author}
  {\bibfnamefont {A.~H.}\ \bibnamefont {Safavi-Naeini}},\ }\href@noop {}
  {\bibfield  {journal} {\bibinfo  {journal} {Physical Review Applied}\
  }\textbf {\bibinfo {volume} {15}},\ \bibinfo {pages} {014039} (\bibinfo
  {year} {2021})}\BibitemShut {NoStop}%
\bibitem [{\citenamefont {Ding}\ \emph {et~al.}(2024)\citenamefont {Ding},
  \citenamefont {Pingault}, \citenamefont {Shao}, \citenamefont {Sinclair},
  \citenamefont {Machielse}, \citenamefont {Chia}, \citenamefont {Maity},\ and\
  \citenamefont {Lon{\v{c}}ar}}]{ding2024integrated}%
  \BibitemOpen
  \bibfield  {author} {\bibinfo {author} {\bibfnamefont {S.~W.}\ \bibnamefont
  {Ding}}, \bibinfo {author} {\bibfnamefont {B.}~\bibnamefont {Pingault}},
  \bibinfo {author} {\bibfnamefont {L.}~\bibnamefont {Shao}}, \bibinfo {author}
  {\bibfnamefont {N.}~\bibnamefont {Sinclair}}, \bibinfo {author}
  {\bibfnamefont {B.}~\bibnamefont {Machielse}}, \bibinfo {author}
  {\bibfnamefont {C.}~\bibnamefont {Chia}}, \bibinfo {author} {\bibfnamefont
  {S.}~\bibnamefont {Maity}}, \ and\ \bibinfo {author} {\bibfnamefont
  {M.}~\bibnamefont {Lon{\v{c}}ar}},\ }\href@noop {} {\bibfield  {journal}
  {\bibinfo  {journal} {Physical Review Applied}\ }\textbf {\bibinfo {volume}
  {21}},\ \bibinfo {pages} {014034} (\bibinfo {year} {2024})}\BibitemShut
  {NoStop}%
\bibitem [{\citenamefont {Deng}\ \emph {et~al.}(2025)\citenamefont {Deng},
  \citenamefont {Anderson}, \citenamefont {Du}, \citenamefont {Roberts},
  \citenamefont {Miller}, \citenamefont {Smith}, \citenamefont {Hackett},
  \citenamefont {Olsson},\ and\ \citenamefont
  {Eichenfield}}]{deng2025strongly}%
  \BibitemOpen
  \bibfield  {author} {\bibinfo {author} {\bibfnamefont {Y.}~\bibnamefont
  {Deng}}, \bibinfo {author} {\bibfnamefont {D.}~\bibnamefont {Anderson}},
  \bibinfo {author} {\bibfnamefont {X.}~\bibnamefont {Du}}, \bibinfo {author}
  {\bibfnamefont {W.}~\bibnamefont {Roberts}}, \bibinfo {author} {\bibfnamefont
  {M.}~\bibnamefont {Miller}}, \bibinfo {author} {\bibfnamefont
  {B.}~\bibnamefont {Smith}}, \bibinfo {author} {\bibfnamefont
  {L.}~\bibnamefont {Hackett}}, \bibinfo {author} {\bibfnamefont
  {T.}~\bibnamefont {Olsson}}, \ and\ \bibinfo {author} {\bibfnamefont
  {M.}~\bibnamefont {Eichenfield}},\ }\href@noop {} {\bibfield  {journal}
  {\bibinfo  {journal} {arXiv preprint arXiv:2503.18113}\ } (\bibinfo {year}
  {2025})}\BibitemShut {NoStop}%
\bibitem [{\citenamefont {Atatüre}\ \emph {et~al.}(2018)\citenamefont
  {Atatüre}, \citenamefont {Englund}, \citenamefont {Vamivakas}, \citenamefont
  {Lee},\ and\ \citenamefont {Wrachtrup}}]{atature_material_2018}%
  \BibitemOpen
  \bibfield  {author} {\bibinfo {author} {\bibfnamefont {M.}~\bibnamefont
  {Atatüre}}, \bibinfo {author} {\bibfnamefont {D.}~\bibnamefont {Englund}},
  \bibinfo {author} {\bibfnamefont {N.}~\bibnamefont {Vamivakas}}, \bibinfo
  {author} {\bibfnamefont {S.-Y.}\ \bibnamefont {Lee}}, \ and\ \bibinfo
  {author} {\bibfnamefont {J.}~\bibnamefont {Wrachtrup}},\ }\href {\doibase
  10.1038/s41578-018-0008-9} {\bibfield  {journal} {\bibinfo  {journal} {Nature
  Reviews Materials}\ }\textbf {\bibinfo {volume} {3}},\ \bibinfo {pages} {38}
  (\bibinfo {year} {2018})}\BibitemShut {NoStop}%
\bibitem [{\citenamefont {Meesala}\ \emph {et~al.}(2018)\citenamefont
  {Meesala}, \citenamefont {Sohn}, \citenamefont {Pingault}, \citenamefont
  {Shao}, \citenamefont {Atikian}, \citenamefont {Holzgrafe}, \citenamefont
  {G{\"u}ndo{\u{g}}an}, \citenamefont {Stavrakas}, \citenamefont {Sipahigil},
  \citenamefont {Chia} \emph {et~al.}}]{meesala2018strain}%
  \BibitemOpen
  \bibfield  {author} {\bibinfo {author} {\bibfnamefont {S.}~\bibnamefont
  {Meesala}}, \bibinfo {author} {\bibfnamefont {Y.-I.}\ \bibnamefont {Sohn}},
  \bibinfo {author} {\bibfnamefont {B.}~\bibnamefont {Pingault}}, \bibinfo
  {author} {\bibfnamefont {L.}~\bibnamefont {Shao}}, \bibinfo {author}
  {\bibfnamefont {H.~A.}\ \bibnamefont {Atikian}}, \bibinfo {author}
  {\bibfnamefont {J.}~\bibnamefont {Holzgrafe}}, \bibinfo {author}
  {\bibfnamefont {M.}~\bibnamefont {G{\"u}ndo{\u{g}}an}}, \bibinfo {author}
  {\bibfnamefont {C.}~\bibnamefont {Stavrakas}}, \bibinfo {author}
  {\bibfnamefont {A.}~\bibnamefont {Sipahigil}}, \bibinfo {author}
  {\bibfnamefont {C.}~\bibnamefont {Chia}},  \emph {et~al.},\ }\href@noop {}
  {\bibfield  {journal} {\bibinfo  {journal} {Physical Review B}\ }\textbf
  {\bibinfo {volume} {97}},\ \bibinfo {pages} {205444} (\bibinfo {year}
  {2018})}\BibitemShut {NoStop}%
\bibitem [{\citenamefont {Maity}\ \emph {et~al.}(2020)\citenamefont {Maity},
  \citenamefont {Shao}, \citenamefont {Bogdanovi{\'c}}, \citenamefont
  {Meesala}, \citenamefont {Sohn}, \citenamefont {Sinclair}, \citenamefont
  {Pingault}, \citenamefont {Chalupnik}, \citenamefont {Chia}, \citenamefont
  {Zheng} \emph {et~al.}}]{maity2020coherent}%
  \BibitemOpen
  \bibfield  {author} {\bibinfo {author} {\bibfnamefont {S.}~\bibnamefont
  {Maity}}, \bibinfo {author} {\bibfnamefont {L.}~\bibnamefont {Shao}},
  \bibinfo {author} {\bibfnamefont {S.}~\bibnamefont {Bogdanovi{\'c}}},
  \bibinfo {author} {\bibfnamefont {S.}~\bibnamefont {Meesala}}, \bibinfo
  {author} {\bibfnamefont {Y.-I.}\ \bibnamefont {Sohn}}, \bibinfo {author}
  {\bibfnamefont {N.}~\bibnamefont {Sinclair}}, \bibinfo {author}
  {\bibfnamefont {B.}~\bibnamefont {Pingault}}, \bibinfo {author}
  {\bibfnamefont {M.}~\bibnamefont {Chalupnik}}, \bibinfo {author}
  {\bibfnamefont {C.}~\bibnamefont {Chia}}, \bibinfo {author} {\bibfnamefont
  {L.}~\bibnamefont {Zheng}},  \emph {et~al.},\ }\href@noop {} {\bibfield
  {journal} {\bibinfo  {journal} {Nature communications}\ }\textbf {\bibinfo
  {volume} {11}},\ \bibinfo {pages} {193} (\bibinfo {year} {2020})}\BibitemShut
  {NoStop}%
\bibitem [{\citenamefont {Sukachev}\ \emph
  {et~al.}(2017{\natexlab{a}})\citenamefont {Sukachev}, \citenamefont
  {Sipahigil}, \citenamefont {Nguyen}, \citenamefont {Bhaskar}, \citenamefont
  {Evans}, \citenamefont {Jelezko},\ and\ \citenamefont
  {Lukin}}]{sukachev2017silicon}%
  \BibitemOpen
  \bibfield  {author} {\bibinfo {author} {\bibfnamefont {D.~D.}\ \bibnamefont
  {Sukachev}}, \bibinfo {author} {\bibfnamefont {A.}~\bibnamefont {Sipahigil}},
  \bibinfo {author} {\bibfnamefont {C.~T.}\ \bibnamefont {Nguyen}}, \bibinfo
  {author} {\bibfnamefont {M.~K.}\ \bibnamefont {Bhaskar}}, \bibinfo {author}
  {\bibfnamefont {R.~E.}\ \bibnamefont {Evans}}, \bibinfo {author}
  {\bibfnamefont {F.}~\bibnamefont {Jelezko}}, \ and\ \bibinfo {author}
  {\bibfnamefont {M.~D.}\ \bibnamefont {Lukin}},\ }\href@noop {} {\bibfield
  {journal} {\bibinfo  {journal} {Physical review letters}\ }\textbf {\bibinfo
  {volume} {119}},\ \bibinfo {pages} {223602} (\bibinfo {year}
  {2017}{\natexlab{a}})}\BibitemShut {NoStop}%
\bibitem [{\citenamefont {Lemonde}\ \emph {et~al.}(2018)\citenamefont
  {Lemonde}, \citenamefont {Meesala}, \citenamefont {Sipahigil}, \citenamefont
  {Schuetz}, \citenamefont {Lukin}, \citenamefont {Loncar},\ and\ \citenamefont
  {Rabl}}]{lemonde2018phonon}%
  \BibitemOpen
  \bibfield  {author} {\bibinfo {author} {\bibfnamefont {M.-A.}\ \bibnamefont
  {Lemonde}}, \bibinfo {author} {\bibfnamefont {S.}~\bibnamefont {Meesala}},
  \bibinfo {author} {\bibfnamefont {A.}~\bibnamefont {Sipahigil}}, \bibinfo
  {author} {\bibfnamefont {M.}~\bibnamefont {Schuetz}}, \bibinfo {author}
  {\bibfnamefont {M.}~\bibnamefont {Lukin}}, \bibinfo {author} {\bibfnamefont
  {M.}~\bibnamefont {Loncar}}, \ and\ \bibinfo {author} {\bibfnamefont
  {P.}~\bibnamefont {Rabl}},\ }\href@noop {} {\bibfield  {journal} {\bibinfo
  {journal} {Physical review letters}\ }\textbf {\bibinfo {volume} {120}},\
  \bibinfo {pages} {213603} (\bibinfo {year} {2018})}\BibitemShut {NoStop}%
\bibitem [{\citenamefont {Yamamoto}\ \emph {et~al.}(2023)\citenamefont
  {Yamamoto}, \citenamefont {Kurokawa}, \citenamefont {Fujii}, \citenamefont
  {Makino}, \citenamefont {Kato},\ and\ \citenamefont
  {Kosaka}}]{yamamoto2023low}%
  \BibitemOpen
  \bibfield  {author} {\bibinfo {author} {\bibfnamefont {M.}~\bibnamefont
  {Yamamoto}}, \bibinfo {author} {\bibfnamefont {H.}~\bibnamefont {Kurokawa}},
  \bibinfo {author} {\bibfnamefont {S.}~\bibnamefont {Fujii}}, \bibinfo
  {author} {\bibfnamefont {T.}~\bibnamefont {Makino}}, \bibinfo {author}
  {\bibfnamefont {H.}~\bibnamefont {Kato}}, \ and\ \bibinfo {author}
  {\bibfnamefont {H.}~\bibnamefont {Kosaka}},\ }\href@noop {} {\bibfield
  {journal} {\bibinfo  {journal} {Journal of Applied Physics}\ }\textbf
  {\bibinfo {volume} {134}} (\bibinfo {year} {2023})}\BibitemShut {NoStop}%
\bibitem [{\citenamefont {Dahmani}\ \emph {et~al.}(2020)\citenamefont
  {Dahmani}, \citenamefont {Sarabalis}, \citenamefont {Jiang}, \citenamefont
  {Mayor},\ and\ \citenamefont {Safavi-Naeini}}]{dahmani2020piezoelectric}%
  \BibitemOpen
  \bibfield  {author} {\bibinfo {author} {\bibfnamefont {Y.~D.}\ \bibnamefont
  {Dahmani}}, \bibinfo {author} {\bibfnamefont {C.~J.}\ \bibnamefont
  {Sarabalis}}, \bibinfo {author} {\bibfnamefont {W.}~\bibnamefont {Jiang}},
  \bibinfo {author} {\bibfnamefont {F.~M.}\ \bibnamefont {Mayor}}, \ and\
  \bibinfo {author} {\bibfnamefont {A.~H.}\ \bibnamefont {Safavi-Naeini}},\
  }\href@noop {} {\bibfield  {journal} {\bibinfo  {journal} {Physical Review
  Applied}\ }\textbf {\bibinfo {volume} {13}},\ \bibinfo {pages} {024069}
  (\bibinfo {year} {2020})}\BibitemShut {NoStop}%
\bibitem [{\citenamefont {Sarabalis}\ \emph {et~al.}(2020)\citenamefont
  {Sarabalis}, \citenamefont {Dahmani}, \citenamefont {Cleland},\ and\
  \citenamefont {Safavi-Naeini}}]{sarabalis_s-band_2020}%
  \BibitemOpen
  \bibfield  {author} {\bibinfo {author} {\bibfnamefont {C.~J.}\ \bibnamefont
  {Sarabalis}}, \bibinfo {author} {\bibfnamefont {Y.~D.}\ \bibnamefont
  {Dahmani}}, \bibinfo {author} {\bibfnamefont {A.~Y.}\ \bibnamefont
  {Cleland}}, \ and\ \bibinfo {author} {\bibfnamefont {A.~H.}\ \bibnamefont
  {Safavi-Naeini}},\ }\href {\doibase 10.1063/1.5126428} {\bibfield  {journal}
  {\bibinfo  {journal} {Journal of Applied Physics}\ }\textbf {\bibinfo
  {volume} {127}},\ \bibinfo {pages} {054501} (\bibinfo {year}
  {2020})}\BibitemShut {NoStop}%
\bibitem [{\citenamefont {Meitl}\ \emph {et~al.}(2006)\citenamefont {Meitl},
  \citenamefont {Zhu}, \citenamefont {Kumar}, \citenamefont {Lee},
  \citenamefont {Feng}, \citenamefont {Huang}, \citenamefont {Adesida},
  \citenamefont {Nuzzo},\ and\ \citenamefont {Rogers}}]{meitl2006transfer}%
  \BibitemOpen
  \bibfield  {author} {\bibinfo {author} {\bibfnamefont {M.~A.}\ \bibnamefont
  {Meitl}}, \bibinfo {author} {\bibfnamefont {Z.-T.}\ \bibnamefont {Zhu}},
  \bibinfo {author} {\bibfnamefont {V.}~\bibnamefont {Kumar}}, \bibinfo
  {author} {\bibfnamefont {K.~J.}\ \bibnamefont {Lee}}, \bibinfo {author}
  {\bibfnamefont {X.}~\bibnamefont {Feng}}, \bibinfo {author} {\bibfnamefont
  {Y.~Y.}\ \bibnamefont {Huang}}, \bibinfo {author} {\bibfnamefont
  {I.}~\bibnamefont {Adesida}}, \bibinfo {author} {\bibfnamefont {R.~G.}\
  \bibnamefont {Nuzzo}}, \ and\ \bibinfo {author} {\bibfnamefont {J.~A.}\
  \bibnamefont {Rogers}},\ }\href@noop {} {\bibfield  {journal} {\bibinfo
  {journal} {Nature Materials}\ }\textbf {\bibinfo {volume} {5}},\ \bibinfo
  {pages} {33} (\bibinfo {year} {2006})}\BibitemShut {NoStop}%
\bibitem [{\citenamefont {Jiang}\ \emph {et~al.}(2023)\citenamefont {Jiang},
  \citenamefont {Mayor}, \citenamefont {Malik}, \citenamefont {Van~Laer},
  \citenamefont {McKenna}, \citenamefont {Patel}, \citenamefont {Witmer},\ and\
  \citenamefont {Safavi-Naeini}}]{jiang2023optically}%
  \BibitemOpen
  \bibfield  {author} {\bibinfo {author} {\bibfnamefont {W.}~\bibnamefont
  {Jiang}}, \bibinfo {author} {\bibfnamefont {F.~M.}\ \bibnamefont {Mayor}},
  \bibinfo {author} {\bibfnamefont {S.}~\bibnamefont {Malik}}, \bibinfo
  {author} {\bibfnamefont {R.}~\bibnamefont {Van~Laer}}, \bibinfo {author}
  {\bibfnamefont {T.~P.}\ \bibnamefont {McKenna}}, \bibinfo {author}
  {\bibfnamefont {R.~N.}\ \bibnamefont {Patel}}, \bibinfo {author}
  {\bibfnamefont {J.~D.}\ \bibnamefont {Witmer}}, \ and\ \bibinfo {author}
  {\bibfnamefont {A.~H.}\ \bibnamefont {Safavi-Naeini}},\ }\href@noop {}
  {\bibfield  {journal} {\bibinfo  {journal} {Nature Physics}\ }\textbf
  {\bibinfo {volume} {19}},\ \bibinfo {pages} {1423} (\bibinfo {year}
  {2023})}\BibitemShut {NoStop}%
\bibitem [{\citenamefont {Hepp}\ \emph {et~al.}(2014)\citenamefont {Hepp},
  \citenamefont {Müller}, \citenamefont {Waselowski}, \citenamefont {Becker},
  \citenamefont {Pingault}, \citenamefont {Sternschulte}, \citenamefont
  {Steinmüller-Nethl}, \citenamefont {Gali}, \citenamefont {Maze},
  \citenamefont {Atatüre},\ and\ \citenamefont
  {Becher}}]{hepp_electronic_2014}%
  \BibitemOpen
  \bibfield  {author} {\bibinfo {author} {\bibfnamefont {C.}~\bibnamefont
  {Hepp}}, \bibinfo {author} {\bibfnamefont {T.}~\bibnamefont {Müller}},
  \bibinfo {author} {\bibfnamefont {V.}~\bibnamefont {Waselowski}}, \bibinfo
  {author} {\bibfnamefont {J.~N.}\ \bibnamefont {Becker}}, \bibinfo {author}
  {\bibfnamefont {B.}~\bibnamefont {Pingault}}, \bibinfo {author}
  {\bibfnamefont {H.}~\bibnamefont {Sternschulte}}, \bibinfo {author}
  {\bibfnamefont {D.}~\bibnamefont {Steinmüller-Nethl}}, \bibinfo {author}
  {\bibfnamefont {A.}~\bibnamefont {Gali}}, \bibinfo {author} {\bibfnamefont
  {J.~R.}\ \bibnamefont {Maze}}, \bibinfo {author} {\bibfnamefont
  {M.}~\bibnamefont {Atatüre}}, \ and\ \bibinfo {author} {\bibfnamefont
  {C.}~\bibnamefont {Becher}},\ }\href {\doibase
  10.1103/PhysRevLett.112.036405} {\bibfield  {journal} {\bibinfo  {journal}
  {Physical Review Letters}\ }\textbf {\bibinfo {volume} {112}},\ \bibinfo
  {pages} {036405} (\bibinfo {year} {2014})}\BibitemShut {NoStop}%
\bibitem [{\citenamefont {Sukachev}\ \emph
  {et~al.}(2017{\natexlab{b}})\citenamefont {Sukachev}, \citenamefont
  {Sipahigil}, \citenamefont {Nguyen}, \citenamefont {Bhaskar}, \citenamefont
  {Evans}, \citenamefont {Jelezko},\ and\ \citenamefont
  {Lukin}}]{sukachev_silicon-vacancy_2017}%
  \BibitemOpen
  \bibfield  {author} {\bibinfo {author} {\bibfnamefont {D.}~\bibnamefont
  {Sukachev}}, \bibinfo {author} {\bibfnamefont {A.}~\bibnamefont {Sipahigil}},
  \bibinfo {author} {\bibfnamefont {C.}~\bibnamefont {Nguyen}}, \bibinfo
  {author} {\bibfnamefont {M.}~\bibnamefont {Bhaskar}}, \bibinfo {author}
  {\bibfnamefont {R.}~\bibnamefont {Evans}}, \bibinfo {author} {\bibfnamefont
  {F.}~\bibnamefont {Jelezko}}, \ and\ \bibinfo {author} {\bibfnamefont
  {M.}~\bibnamefont {Lukin}},\ }\href {\doibase 10.1103/PhysRevLett.119.223602}
  {\bibfield  {journal} {\bibinfo  {journal} {Physical Review Letters}\
  }\textbf {\bibinfo {volume} {119}},\ \bibinfo {pages} {223602} (\bibinfo
  {year} {2017}{\natexlab{b}})}\BibitemShut {NoStop}%
\end{thebibliography}%

\end{document}